\documentstyle[12pt]{article}

\begin{document}

\begin{center}
Continuum Field Model of Street Canyon: Numerical Examples
\end{center}
\begin{center}
Part 2
\end{center}
\begin{center}
Maciej M. Duras $\star$
\end{center}
\begin{center}
Institute of Physics, Cracow University of Technology, 
ulica Podchor\c{a}\.zych 1, PL-30-084 Cracow, Poland
\end{center}
\begin{center}
$\star$ Email: mduras @ riad.usk.pk.edu.pl
\end{center}
\begin{center}
{\em Engng. Trans.}  {\bf 48}, 433-448 (2000). 
\end{center}

\begin{abstract}
Continuum field control model of a street canyon is considered.
The six separate monocriterial optimal control problems 
consist of minimization of functionals 
of the total travel time,
of global emissions of pollutants, 
and of global concentrations of pollutants,
both in the studied street canyon,
and in its two nearest neighbour substitute canyons, respectively.
The six optimization problems for the functionals
are solved numerically.
General traffic control issues are inferred.
The discretization method, comparison with experiment,
mathematical issues, and programming issues, are
presented.
\end{abstract}

\section{Introduction.}
\label{sect-introduction}
\setcounter{equation}{0}  
The present article is a continuation of \cite{Duras 1999 engtrans theory}.
The notation of \cite{Duras 1999 engtrans theory} will be used.
Now, we will solve numerically six separate monocriterial
optimization problems {\bf O1-O6}.
We assumed the data
from real street canyon of the Krasi\'nski Avenue 
in Cracow \cite{Brzezanski 1998}.
We solve the set of nonlinear 
partial differential equations {\bf E1-E8} with given 
boundary {\bf B0-B8}, and the initial conditions {\bf C0-C8},
and sources {\bf D0-D2}, by the finite difference method, 
using the C language programme written by the author. 
We solve this set in the cuboid $\Omega$, starting from initial 
conditions, and we iterate it over the time period  
using the direct finite difference method 
taking into account the boundary {\bf B0-B8} conditions,
and the sources {\bf D0-D2}, and initial  
conditions {\bf C0-C8}, at each time step. 
The functionals in {\bf F1-F6} 
are iterated with the same steps as the equations {\bf E1-E8}
are iterated.
The first time derivative is approximated by
the first differential quotient 
using forward two-point first difference 
in the direction of time coordinate, 
whereas the spatial first derivatives are approximated by
the first differential quotients 
using central three-point first differences in the directions of  
spatial coordinates. 
In general, the numerical results
are in good agreement 
with the measured data from \cite{Brzezanski 1998}.

\section{General Traffic Management Inferences.}
\label{sect-gen-traff-illat}
\setcounter{equation}{0}  
On the basis of the performed simulations, we deduce the traffic
management inferences.
In general, some vehicular traffic and hydrodynamical parameters
influence the solutions of optimization problems
{\bf O1-O6}, but not all of them.

{\bf R1.} The direction of velocity of air mixture is important.
The optimal pollutant concentrations for both the single canyon
and the canyon with its two substitute nearest neighbour canyons,
{\bf O3, O6},
are the lowest ones if the velocity components 
of the boundary and the initial value problems
are equal to zero;
further, they are grater 
if the velocity has only nonzero vertical component $v_{z}$;
then, they are greater if velocity has only nonzero x-component $v_{x}$;
further, they are again greater if the velocity has two nonzero components
$(0, v_{y}, v_{z})$;
next, they are again greater if the velocity has three nonzero components
$(v_{x}, v_{y}, v_{z})$;
finally, they are the greatest if the velocity has only 
nonzero y-component $v_{y}$.
We infer that in case R1.6,
the optimal green time $g_{2, {\rm C}}^{\star}$ is different from 
other cases. It means that velocity of the mixture
influences optimal vectors of control.

{\bf R2.} The optimal values for cases {\bf O1, O4}, 
for cases {\bf O2, O5}, and for cases {\bf O3, O6}, 
decrease with ``uniformization'' of the vehicles, 
when we pass from nonuniform vehicles ${\rm UNIFORM}=0$ 
to uniform ones ${\rm UNIFORM}=1$. 
It is a result of reduction of the number of vehicles moving in the canyon. 
For ${\rm UNIFORM}=1$ the values of maximum free flow speed, 
jam, saturation, threshold, green, and red densities 
reach the minimal values
\cite{Duras 1998 thesis}.

{\bf R3.} The long vehicular queues decrease 
the total travel times {\bf O1, O4},
and they increase both the optimal emissions {\bf O2, O5},
and optimal concentrations of pollutants {\bf O3, O6}
\cite{Duras 1998 thesis}.
The decrement of total travel times {\bf O1, O4},
with long vehicular queues is a result of clustering of the vehicles.

{\bf R4.} The constant of temperature scale ${\rm TH}$ 
does not differentiate the values 
of optimal concentrations of pollutants {\bf O3, O6}
in the temperature range near standard temperature and pressure STP 
conditions. However, it diminishes them even hundredfold
for very high temperatures .

{\bf R5.} The functional form of initial and boundary conditions
affects the optima. If they are constant, then optimal concentrations
{\bf O3, O6}, are two times higher than in the case when they
are changing exponentially in space in three dimensions. 
The index ${\rm FORM=CONSTANT}$ corresponds
to constant initial and boundary conditions,
whereas ${\rm FORM=EXPONENTIAL}$ corresponds
to the spatially exponentially changing initial and boundary conditions.
For the latter case, 
we chose the following functional form of initial conditions:
the temperature is equal to 
$T_{0}(x,y,z,t)=\exp(-\frac{x}{a}-\frac{y}{b}-\frac{z}{c}) {\rm TH}$,
the only non-zero coordinates of velocities of mixture are the  
x-coordinates and they are equal to 
$v_{x,0}(x,y,z,t)=\exp(-\frac{x}{a}-\frac{y}{b}-\frac{z}{c}) {\rm VX}$,
the density is equal to 
$\rho_{0}(x,y,z,t)=\exp(-\frac{x}{a}-\frac{y}{b}-\frac{z}{c}) \rho_{{\rm STP}}$,
the concentrations are equal to 
$c_{i,0}(x,y,z,t)=\exp(-\frac{x}{a}-\frac{y}{b}-\frac{z}{c}) c_{i, {\rm STP}}$,
and the pressure is equal to
$p_{0}(x,y,z,t)=\exp(-\frac{x}{a}-\frac{y}{b}-\frac{z}{c}) p_{{\rm STP}}$.
The boundary conditions {\bf B0-B8} are similar.

{\bf R6.} The presence of vehicles on both the left and
right lanes is important.
The optimal total travel times and emissions are halved in absence of
vehicles on the left or right lanes
when compared to situation when they circulate on both the left and
right lanes
\cite{Duras 1998 thesis}.

{\bf R7.} The values of saturation, arrival, or jam vehicular density,
and of vehicular free flow velocities also affect the optima
{\bf O1-O6}
\cite{Duras 1998 thesis}.

{\bf R8.} The assumption of energy conservation equation,
of the thermodiffusion effect,
of the chemical potential and of the Grand Canonical ensemble,
as well as of influence of gravity on intrinsic energy and
on the chemical potential,
drastically changes the optimal concentrations {\bf O3, O6},
towards the measured ones 
\cite{Duras 1998 thesis, Duras 1997 PJES, Duras 1999 PJES}. 

{\bf R9.} The value of time of simulation and of
discretization in time considerably affects the optima
(compare \cite{Duras 1998 thesis}). 
The values of optimal solutions {\bf O1-O6},
increase from tenfold to hundredfold.
Also the optimal vectors (5-tuples) of control for {\bf O1-O6},
change their values.
It is a result of cumulative effect 
of length of the period of simulation $T_{S}$
on integral functionals {\bf F1-F6}. 

We use the following notation: 
${\rm yL}=0/{\rm yR}=0$ means that the lengths of all queues 
on the left/right lanes at the beginning were equal to zero, 
${\rm VX}=0$ is meant for zero initial and boundary velocity of 
mixture, while  ${\rm VX}=1$ is put 
for the velocity of ${\rm 1 [m \cdot s^{-1}]}$. 
The same holds for ${\rm VY}$ and ${\rm VZ}$.
If ${\rm VX} \neq 0$, then the left lanes 
are leeward and the right ones windward. 
If there are no vehicles on the left or right lanes, then 
${\rm LON}=0$ or ${\rm RON}=0$, respectively.
${\rm UNIFORM}=0$ stands for non-homogeneous (different)
values of maximum free flow speed $w_{vt, f}$, jam $k_{vt, {\rm jam}}$, 
saturation $k_{vt, {\rm sat}}$, threshold $k_{vt, {\rm threshold}}$,
green $k_{vt, {\rm GREEN}}$, 
and red $k_{vt, {\rm RED}}$, vehicular densities  
for $VT=4$ types of vehicles: passenger cars, 
8-ton, 12-ton, and 16-ton trucks,
where threshold vehicular densities 
$k_{vt, {\rm threshold}}$ are the numerical values
for which we assume in numerical simulations that there is
a vehicle on the given lane.
We also assume that these values are the same for both the left
and right lanes. 
$\delta_{x}, \delta_{y}, \delta_{z}, \delta_{t}$,
are the steps in $x, y, z, t,$ directions, respectively.
$\delta_{C_{1}}, \delta_{g_{1}}, \delta_{C_{2}}, \delta_{g_{2}}, \delta_{F}$,
are the steps in
$C_{1}, g_{1}, C_{2}, g_{2}, F$, directions, respectively.
In the boundary and initial conditions {\bf B0a, B0b, B0c, C0}, 
we assume that the temperature is equal to a given constant ${\rm TH}$.  

\section{Discretization scheme.}
\label{sect-discretization}
\setcounter{equation}{0}  

In order to solve numerically the set of nonlinear partial 
differential equations {\bf E1}-{\bf E8}, one uses the apparatus
of finite differences. Firstly, we discretize the points of domain $\Sigma$
in the standard way:
\begin{equation}
(x, y, z, t) \simeq (x_{i}, y_{j}, z_{k}, t_{l}),
i=0, \cdots, N_{x},j=0, \cdots, N_{y},k=0, \cdots, N_{z},l=0, \cdots, N_{t},
\label{xyzt-ijkl}
\end{equation}

\begin{equation}
x_{i}=x_{0}+\frac{i-1}{N_{x}}(x_{N_{x}}-x_{0}),
\label{x-i-def}
\end{equation}

\begin{equation}
y_{i}=y_{0}+\frac{j-1}{N_{y}}(y_{N_{y}}-y_{0}),
\label{y-j-def}
\end{equation}

\begin{equation}
z_{k}=z_{0}+\frac{k-1}{N_{z}}(z_{N_{z}}-z_{0}),
\label{z-k-def}
\end{equation}

\begin{equation}
t_{l}=t_{0}+\frac{l-1}{N_{t}}(t_{N_{t}}-t_{0}).
\label{t-l-def}
\end{equation}
Secondly, the value of given function $f$ at point $(x, y, z, t)$ 
of its domain $\Sigma$ is approximated by:
\begin{equation}
f(x, y, z, t) \simeq f(x_{i}, y_{i}, z_{k}, t_{l}) \equiv
f_{i,j,k,l}.
\label{xyzt-ijkl-function}
\end{equation}
Thirdly, the first finite differences in spatial and temporal directions:
\begin{eqnarray}
& & \Delta_{x}f_{i,j,k,l}=\frac{1}{2}(f_{i+1,j,k,l}-f_{i-1,j,k,l}),
\label{xyzt-ijkl-finite-differences}
\\
& & \Delta_{y}f_{i,j,k,l}=\frac{1}{2}(f_{i,j+1,k,l}-f_{i,j-1,k,l}),
\\
& & \Delta_{z}f_{i,j,k,l}=\frac{1}{2}(f_{i,j,k+1,l}-f_{i,j,k-1,l}),
\\
& & \Delta_{t}f_{i,j,k,l}=f_{i,j,k,l+1}-f_{i,j,k,l},
\end{eqnarray}
allow to approximate the first partial derivatives:
\begin{eqnarray}
& & \frac{\partial f(x, y, z, t)}{\partial x} 
\simeq \frac{\Delta_{x}f_{i,j,k,l}}{\Delta x},
\label{xyzt-ijkl-derivatives}
\\
& & \frac{\partial f(x, y, z, t)}{\partial y} 
\simeq \frac{\Delta_{y}f_{i,j,k,l}}{\Delta y},
\\
& & \frac{\partial f(x, y, z, t)}{\partial z} 
\simeq \frac{\Delta_{z}f_{i,j,k,l}}{\Delta z},
\\
& & \frac{\partial f(x, y, z, t)}{\partial t} 
\simeq \frac{\Delta_{t}f_{i,j,k,l}}{\Delta t}.
\end{eqnarray}
Since the vector and scalar
partial differential equations {\bf E1}-{\bf E8}
are of the following form:
\begin{equation}
\frac{\partial f^{\alpha}(x, y, z, t)}{\partial t} =
F^{\alpha}(x, y, z, t, f^{\alpha}(x, y, z, t), 
\frac{\partial f^{\alpha}(x, y, z, t)}{\partial x_{m}}, 
\frac{\partial^{2} f^{\alpha}(x, y, z, t)}{\partial x_{m} \partial x_{n}},
f^{\beta}(x, y, z, t)),
\label{xyzt-ijkl-pde}
\end{equation} 
hence we approximate them as follows:
\begin{equation}
\frac{\Delta_{t} f_{i,j,k,l}^{\alpha}}{\Delta t} =
F^{\alpha}(x_{i}, y_{j}, z_{k}, t_{l},
f_{i,j,k,l}^{\alpha}, \frac{\Delta_{x_{m}} f_{i,j,k,l}^{\alpha}}{\Delta x_{m}},
\frac{\Delta_{x_{m}} \Delta_{x_{n}} f_{i,j,k,l}^{\alpha}}
{\Delta x_{m} \Delta x_{n}},
f_{i,j,k,l}^{\beta}).
\label{xyzt-ijkl-discretization}
\end{equation}
Hence, we can rewrite it in following forward scheme:
\begin{equation}
f_{i,j,k,l+1}^{\alpha}=f_{i,j,k,l}^{\alpha} + \Delta t \cdot
F^{\alpha}(x_{i}, y_{j}, z_{k}, t_{l},
f_{i,j,k,l}^{\alpha}, \frac{\Delta_{x_{m}} f_{i,j,k,l}^{\alpha}}{\Delta x_{m}},
\frac{\Delta_{x_{m}} \Delta_{x_{n}} f_{i,j,k,l}^{\alpha}}
{\Delta x_{m} \Delta x_{n}},
f_{i,j,k,l}^{\beta}).
\label{xyzt-ijkl-solution}
\end{equation}

We give now the examples of dicretized equations.
The discretization of continuity equation {\bf E2} reads:
\begin{eqnarray}
& & \rho_{i, j, k, l+1}=
\rho_{i, j, k, l}
- \frac{\Delta t}{2 \Delta x}
v_{i, j, k, l}^{x}
(\rho_{i+1, j, k, l}-\rho_{i-1, j, k, l}) 
\label{xyzt-ijkl-continuity}
\\
& & - \frac{\Delta t}{2 \Delta y}
v_{i, j, k, l}^{y}
(\rho_{i, j+1, k, l}-\rho_{i, j-1, k, l}) 
\nonumber
\\
& & - \frac{\Delta t}{2 \Delta z}
v_{i, j, k, l}^{z}
(\rho_{i, j, k+1, l}-\rho_{i, j, k-1, l}) 
\nonumber
\\
& & + \Delta t \frac{T_{S}}{\rho_{0}} S_{i, j, k, l}
\nonumber
\\
& & - \frac{\Delta t}{2 \Delta x}
\rho_{i, j, k, l}
(\rho_{i+1, j, k, l}-\rho_{i-1, j, k, l}) 
\nonumber
\\
& & - \frac{\Delta t}{2 \Delta y}
\rho_{i, j, k, l}
(\rho_{i, j+1, k, l}-\rho_{i, j-1, k, l}) 
\nonumber
\\
& & - \frac{\Delta t}{2 \Delta z}
\rho_{i, j, k, l}
(\rho_{i, j, k+1, l}-\rho_{i, j, k-1, l}). 
\nonumber
\end{eqnarray}

For the x-component of Navier-Stokes equation {\bf E1} we have:  
\begin{eqnarray}
& & v_{i, j, k, l+1}^{x}=
v_{i, j, k, l}^{x}
- \frac{\Delta t}{2 \Delta x}
v_{i, j, k, l}^{x}
(v_{i+1, j, k, l}^{x}-v_{i-1, j, k, l}^{x}) 
\label{xyzt-ijkl-Navier-Stokes-x}
\\
& & - \frac{\Delta t}{2 \Delta y}
v_{i, j, k, l}^{y}
(v_{i, j+1, k, l}^{x}-v_{i, j-1, k, l}^{x}) 
\nonumber
\\
& & - \frac{\Delta t}{2 \Delta z}
v_{i, j, k, l}^{z}
(v_{i, j, k+1, l}^{x}-v_{i, j, k-1, l}^{x}) 
\nonumber
\\
& & - \Delta t \frac{T_{S}}{\rho_{0}}
\frac{v_{i, j, k, l}^{x}}{\rho_{i, j, k, l}}
S_{i, j, k, l}
\nonumber
\\
& & - \frac{\Delta t}{2 \Delta x}
B T_{0} \frac{T_{S}^{2}}{a^{2}}
(T_{i+1, j, k, l}-T_{i-1, j, k, l}) 
\nonumber
\\
& & - \frac{\Delta t}{2 \Delta x}
B T_{0} \frac{T_{S}^{2}}{a^{2}}
\frac{T_{i, j, k, l}}{\rho_{i, j, k, l}}
(\rho_{i+1, j, k, l}-\rho_{i-1, j, k, l}) 
\nonumber
\\
& & + \frac{\Delta t}{(\Delta x)^2}
\frac{\eta T_{S}}{\rho_{0} a^{2}}
\frac{1}{\rho_{i, j, k, l}}
(v_{i+1, j, k, l}^{x}-2v_{i, j, k, l}^{x}+v_{i-1, j, k, l}^{x}) 
\nonumber
\\
& & + \frac{\Delta t}{(\Delta y)^2}
\frac{\eta T_{S}}{\rho_{0} b^{2}}
\frac{1}{\rho_{i, j, k, l}}
(v_{i, j+1, k, l}^{x}-2v_{i, j, k, l}^{x}+v_{i, j-1, k, l}^{x}) 
\nonumber
\\
& & + \frac{\Delta t}{(\Delta z)^2}
\frac{\eta T_{S}}{\rho_{0} c^{2}}
\frac{1}{\rho_{i, j, k, l}}
(v_{i, j, k+1, l}^{x}-2v_{i, j, k, l}^{x}+v_{i, j, k-1, l}^{x}) 
\nonumber
\\
& & + \frac{\Delta t}{(\Delta x)^2}
(\xi + \frac{\eta}{3})
\frac{\eta T_{S}}{\rho_{0} a^{2}}
\frac{1}{\rho_{i, j, k, l}}
(v_{i+1, j, k, l}^{x}-2v_{i, j, k, l}^{x}+v_{i-1, j, k, l}^{x}) 
\nonumber
\\
& & + \frac{\Delta t}{4 \Delta x \Delta y}
(\xi + \frac{\eta}{3})
\frac{\eta T_{S}}{\rho_{0} b^{2}}
\frac{1}{\rho_{i, j, k, l}}
(v_{i+1, j+1, k, l}^{x}-v_{i+1, j-1, k, l}^{x}
-v_{i-1, j+1, k, l}^{x}+v_{i-1, j-1, k, l}^{x}) 
\nonumber
\\
& & + \frac{\Delta t}{4 \Delta x \Delta z}
(\xi + \frac{\eta}{3})
\frac{\eta T_{S}}{\rho_{0} c^{2}}
\frac{1}{\rho_{i, j, k, l}}
(v_{i+1, j, k+1, l}^{x}-v_{i+1, j, k-1, l}^{x}
-v_{i-1, j, k+1, l}^{x}+v_{i-1, j, k-1, l}^{x}) 
\nonumber
\\
& & + \Delta t \frac{g_{0} T_{S}^{2}}{a} g^{x}.
\nonumber
\end{eqnarray}

Similarly for the y-component:
\begin{eqnarray}
& & v_{i, j, k, l+1}^{y}=
v_{i, j, k, l}^{y}
- \frac{\Delta t}{2 \Delta x}
v_{i, j, k, l}^{x}
(v_{i+1, j, k, l}^{y}-v_{i-1, j, k, l}^{y}) 
\label{xyzt-ijkl-Navier-Stokes-y}
\\
& & - \frac{\Delta t}{2 \Delta y}
v_{i, j, k, l}^{y}
(v_{i, j+1, k, l}^{y}-v_{i, j-1, k, l}^{y}) 
\nonumber
\\
& & - \frac{\Delta t}{2 \Delta z}
v_{i, j, k, l}^{z}
(v_{i, j, k+1, l}^{z}-v_{i, j, k-1, l}^{z}) 
\nonumber
\\
& & - \Delta t \frac{T_{S}}{\rho_{0}}
\frac{v_{i, j, k, l}^{y}}{\rho_{i, j, k, l}}
S_{i, j, k, l}
\nonumber
\\
& & - \frac{\Delta t}{2 \Delta y}
B T_{0} \frac{T_{S}^{2}}{b^{2}}
(T_{i, j+1, k, l}-T_{i, j-1, k, l}) 
\nonumber
\\
& & - \frac{\Delta t}{2 \Delta y}
B T_{0} \frac{T_{S}^{2}}{b^{2}}
\frac{T_{i, j, k, l}}{\rho_{i, j, k, l}}
(\rho_{i, j+1, k, l}-\rho_{i, j-1, k, l}) 
\nonumber
\\
& & + \frac{\Delta t}{(\Delta x)^2}
\frac{\eta T_{S}}{\rho_{0} a^{2}}
\frac{1}{\rho_{i, j, k, l}}
(v_{i+1, j, k, l}^{y}-2v_{i, j, k, l}^{y}+v_{i-1, j, k, l}^{y}) 
\nonumber
\\
& & + \frac{\Delta t}{(\Delta y)^2}
\frac{\eta T_{S}}{\rho_{0} b^{2}}
\frac{1}{\rho_{i, j, k, l}}
(v_{i, j+1, k, l}^{y}-2v_{i, j, k, l}^{y}+v_{i, j-1, k, l}^{y}) 
\nonumber
\\
& & + \frac{\Delta t}{(\Delta z)^2}
\frac{\eta T_{S}}{\rho_{0} c^{2}}
\frac{1}{\rho_{i, j, k, l}}
(v_{i, j, k+1, l}^{y}-2v_{i, j, k, l}^{y}+v_{i, j, k-1, l}^{y}) 
\nonumber
\\
& & + \frac{\Delta t}{4 \Delta x \Delta y}
(\xi + \frac{\eta}{3})
\frac{\eta T_{S}}{\rho_{0} b^{2}}
\frac{1}{\rho_{i, j, k, l}}
(v_{i+1, j+1, k, l}^{x}-v_{i+1, j-1, k, l}^{x}
-v_{i-1, j+1, k, l}^{x}+v_{i-1, j-1, k, l}^{x}) 
\nonumber
\\
& & + \frac{\Delta t}{\Delta y^{2}}
(\xi + \frac{\eta}{3})
\frac{\eta T_{S}}{\rho_{0} b^{2}}
\frac{1}{\rho_{i, j, k, l}}
(v_{i+1, j+1, k, l}^{y}-v_{i+1, j-1, k, l}^{y}
-v_{i-1, j+1, k, l}^{y}+v_{i-1, j-1, k, l}^{y}) 
\nonumber
\\
& & + \frac{\Delta t}{4 \Delta y \Delta z}
(\xi + \frac{\eta}{3})
\frac{\eta T_{S}}{\rho_{0} b^{2}}
\frac{1}{\rho_{i, j, k, l}}
(v_{i, j+1, k+1, l}^{x}-v_{i, j+1, k-1, l}^{x}
-v_{i, j-1, k+1, l}^{x}+v_{i, j-1, k-1, l}^{x}) 
\nonumber
\\
& & + \Delta t \frac{g_{0} T_{S}^{2}}{b} g^{y}.
\nonumber
\end{eqnarray}

Finally, for the z-component: 
\begin{eqnarray}
& & v_{i, j, k, l+1}^{z}=
v_{i, j, k, l}^{z}
- \frac{\Delta t}{2 \Delta x}
v_{i, j, k, l}^{x}
(v_{i+1, j, k, l}^{z}-v_{i-1, j, k, l}^{z}) 
\label{xyzt-ijkl-Navier-Stokes-z}
\\
& & - \frac{\Delta t}{2 \Delta y}
v_{i, j, k, l}^{y}
(v_{i, j+1, k, l}^{z}-v_{i, j-1, k, l}^{z}) 
\nonumber
\\
& & - \frac{\Delta t}{2 \Delta z}
v_{i, j, k, l}^{z}
(v_{i, j, k+1, l}^{z}-v_{i, j, k-1, l}^{z}) 
\nonumber
\\
& & - \Delta t \frac{T_{S}}{\rho_{0}}
\frac{v_{i, j, k, l}^{z}}{\rho_{i, j, k, l}}
S_{i, j, k, l}
\nonumber
\\
& & - \frac{\Delta t}{2 \Delta z}
B T_{0} \frac{T_{S}^{2}}{c^{2}}
(T_{i, j, k+1, l}-T_{i, j, k-1, l}) 
\nonumber
\\
& & - \frac{\Delta t}{2 \Delta z}
B T_{0} \frac{T_{S}^{2}}{c^{2}}
\frac{T_{i, j, k, l}}{\rho_{i, j, k, l}}
(\rho_{i, j, k+1, l}-\rho_{i, j, k-1, l}) 
\nonumber
\\
& & + \frac{\Delta t}{(\Delta x)^2}
\frac{\eta T_{S}}{\rho_{0} a^{2}}
\frac{1}{\rho_{i, j, k, l}}
(v_{i+1, j, k, l}^{z}-2v_{i, j, k, l}^{z}+v_{i-1, j, k, l}^{z}) 
\nonumber
\\
& & + \frac{\Delta t}{(\Delta y)^2}
\frac{\eta T_{S}}{\rho_{0} b^{2}}
\frac{1}{\rho_{i, j, k, l}}
(v_{i, j+1, k, l}^{z}-2v_{i, j, k, l}^{z}+v_{i, j-1, k, l}^{z}) 
\nonumber
\\
& & + \frac{\Delta t}{(\Delta z)^2}
\frac{\eta T_{S}}{\rho_{0} c^{2}}
\frac{1}{\rho_{i, j, k, l}}
(v_{i, j, k+1, l}^{z}-2v_{i, j, k, l}^{z}+v_{i, j, k-1, l}^{z}) 
\nonumber
\\
& & + \frac{\Delta t}{4 \Delta x \Delta z}
(\xi + \frac{\eta}{3})
\frac{\eta T_{S}}{\rho_{0} c^{2}}
\frac{1}{\rho_{i, j, k, l}}
(v_{i+1, j, k+1, l}^{x}-v_{i+1, j, k-1, l}^{x}
-v_{i-1, j, k+1, l}^{x}+v_{i-1, j, k-1, l}^{x}) 
\nonumber
\\
& & + \frac{\Delta t}{4 \Delta y \Delta z}
(\xi + \frac{\eta}{3})
\frac{\eta T_{S}}{\rho_{0} b^{2}}
\frac{1}{\rho_{i, j, k, l}}
(v_{i, j+1, k+1, l}^{x}-v_{i, j, k-1, l}^{x}
-v_{i, j, k+1, l}^{x}+v_{i, j-1, k-1, l}^{x}) 
\nonumber
\\
& & + \frac{\Delta t}{(\Delta z)^{2}}
(\xi + \frac{\eta}{3})
\frac{\eta T_{S}}{\rho_{0} c^{2}}
\frac{1}{\rho_{i, j, k, l}}
(v_{i, j, k+1, l}^{z}-2v_{i, j, k, l}^{z}
+v_{i, j, k-1, l}^{z}) 
\nonumber
\\
& & + \Delta t \frac{g_{0} T_{S}^{2}}{c} g^{z}.
\nonumber
\end{eqnarray} 

\section{Method of comparison.}
\label{sect-method}
\setcounter{equation}{0}  

We briefly describe the method of comparison of the numerical
results with the experimental ones.
The comparison of simulations with the measured data, if authorized,
is aimed at showing the correctness of the model
(conservation of the rank of simulated parameters),
and of presenting the correct direction of deviations
(the deviations always tend to secure the direction). 
We transform both measurements and optima $J_{{\rm C}}^{*}$ {\bf O3},
in order to compare them. 
The measured pollutant concentrations are multiplied by
the volume of the canyon, and we obtain pollutant masses
(we assume that the measured concentrations are the same
in the entire canyon and in time).
The pollutant masses are added up, assuming ${\rm HC}$ mass 
to be equal to zero
(${\rm HC}$ were not measured in Ref. \cite{Brzezanski 1998}).
The calculated optima $J_{{\rm C}}^{*}$ {\bf O3}
are divided by time of simulation $T_{S}$, giving approximation
of pollutant masses in every moment of simulation
(we assume that $J_{{\rm C}}^{*}$ {\bf O3} are homogeneous in time,
in order to make possible the comparison with measurements).
Thence, we have two comparable magnitudes.
Also, only insufficient vehicular flow data are given in
Ref. \cite{Brzezanski 1998}. 
Neither the jam, saturation vehicular densities,
nor the maximum vehicular speeds are present.
However, the initial and boundary conditions of traffic flow
{\bf B5-B8, C5-C8} are not constant functions because
they are parametrized by measured traffic parameters.
The data are not instantaneous,  but they are measured in intervals of hours.
Moreover, the measured data are not accompanied
by measurements of wind velocity, temperature, density.
If the measurements had been performed instantly using
more specialized equipment ({\it e. g.}, lidar),
and if they had been stored in computer facility, 
then, it would be possible to make more serious comparisons.
The simulations were performed for the time period ${\rm T=60 [s]}$,
whereas the measurement were averaged over the time period ${\rm T=1800 [s]}$.
The integral functionals {\bf F3, F6}, are cumulative with respect to
the time period.

\section{Computer simulations.}
\label{sect-computer}
\setcounter{equation}{0}  

From thousands of performed optimizations {\bf O1-O6} 
we selected some representable optimizations.
The C language programme of more than 14000 lines written by the author 
was being run on workstations of Hewlett-Packard, Sun, Silicon Graphics,
under UNIX operating system. It was being run also on 
Pentium personal computers under Linux operating system.
We used C language UNIX/Linux compilers {\tt cc}.
The time of simulations varied from a couple of hours
to several days or even weeks, depending on the discretization
parameters. The discretization steps were tested
in order to obtain finite solutions to the optimization problems
in reasonable time of simulation. They were also studied
from the point of view of sensitiveness of optima appearing on them. 

\section{Mathematical questions.}
\label{sect-mathematical}
\setcounter{equation}{0}  

The problem of uniqueness of solutions
of optimization problems {\bf O1}-{\bf O6} is complex.
If there are no vehicles allowed in the canyon,
then all the six functionals {\bf O1}-{\bf O6}
are constant functions of parameters, the optima exist,
and the number of optimal
solutions is continuum. The values of functionals are then
zero for total travel time and emissions, and are positive
for concentrations.
These results are analytical.
For other cases there are no such analytical results for both
the uniqueness and existence known to the author.  
The numerical solutions of optimization problems
are approximately global within the error induced by
discretization of the physical domain $\Sigma$
and of the domain of control parameters $U^{\rm adm}$.
The method of optimization was a full search.

\section{Conclusions.}
\label{sect-conclusions}
\setcounter{equation}{0}  

The proecological traffic control idea 
and advanced model of the street canyon have been developed. 
It has been found that the proposed model
represents the main features of complex air pollution phenomena.

\section{Acknowledgements.}
\label{sect-acknowledgements}
It is my pleasure to thank Professor Andrzej Adamski
for formulating the problem,
Professors Stanis{\l}aw Bia{\l}as, Wojciech Mitkowski,
and Gwidon Szefer, for many constructive criticisms. 
I also thank Professor W{\l}odzimierz W\'ojcik
for his giving me access to his computer facilities.


\begin{thebibliography}{9}
\bibitem{Duras 1999 engtrans theory}
M. M. DURAS,
Continuum Field Model of Street Canyon: Theoretical Description,
Engineering Transactions, {\bf 48}, 4, 2000.
\bibitem{Brzezanski 1998} 
M. BRZE\.ZA\'NSKI, 
Evaluation of Transport Emission of the Amount 
on a Chosen Section of the Road Network in Cracow, 
KONMOT'98, Table 2, 1998 [in Polish]. 
\bibitem{Duras 1998 thesis}
M. M. DURAS,  
Road Traffic Control in Street Canyons,
Ph. D. Thesis, University of Mining and Metallurgy, Cracow, August 1998, 
pp. 1-363, 1998. 
\bibitem{Duras 1997 PJES} 
A. ADAMSKI, and M. M. DURAS,  
Environmental Traffic Control Issues in Street Canyons,
Polish Journal of Environmental Studies,
{\bf 6}, 1, 13-20, 1997.
\bibitem{Duras 1999 PJES} 
A. ADAMSKI, and M. M. DURAS,  
Air Pollution Optimal Traffic Control Issues in Integrated Street,
Polish Journal of Environmental Studies,
{\bf 8}, 1, 7-17, 1999.
\end{thebibliography}
\end{document}